\documentclass[11pt,twoside
]{article}

\usepackage{baaa2011}
\usepackage{graphicx}
\usepackage{subfigure}
\usepackage{psfrag}
\usepackage{amssymb}
\usepackage[spanish,activeacute,english]{babel}
\usepackage[latin1]{inputenc}
\usepackage[T1]{fontenc} 
\usepackage{ae,aecompl} 
\usepackage{latexsym}
\usepackage{verbatim}
\usepackage{amsmath}
\usepackage{stmaryrd}
\usepackage{amsfonts}
\usepackage{amssymb}
\usepackage{wasysym}
\usepackage[colorlinks=true,dvips]{hyperref}

\begin{document}
\myselectenglish
\vskip 1.0cm
\markboth{ S. Torres Robledo et. al. }%
{Bibliographic compilation of NIR spectroscopy for stars in the GOS Catalog}

\pagestyle{myheadings}
\vspace*{0.5cm}
\noindent PRESENTACI\'ON MURAL
\vskip 0.3cm
\title{Bibliographic compilation of NIR spectroscopy for stars in the Galactic
  O-Star Catalog} 


\author{
 Sim\'on Torres Robledo$^{1}$,  
 Rodolfo Barb\'a$^{1,2}$,
 Julia Arias$^{1}$, \& 
 Nidia Morrell$^{3}$ 
 }

\affil{%
 $^1$ Departamento de F\'isica, Universidad de La  Serena, Av. Cisternas 1200
 Norte, La Serena, Chile\\ 
 $^2$ Instituto de Ciencias Astron\'omicas, de la Tierra y del Espacio,
 CONICET, Casilla 467, 5400 San Juan, Argentina\\ 
 $^3$ Las Campanas Observatory, Observatories of the Carnegie Institution of
 Washington, La Serena, Chile\\
}

\begin{abstract}
We are carrying out a bibliographic compilation of near-infrared (NIR)
($0.7-5.0$ $\mu$m) spectroscopic studies available for stars 
in the Galactic O Star Catalog (GOSC, Ma\'iz Apell\'aniz et al. 2004).
This compilation allows us to quantify the precise degree of knowledge about
NIR spectral information for GOSC sources, such as band coverage, spectral
resolution, equivalent-width measurements, etc.
This bibliographic compilation has a clear next step toward the development of
a new catalog of O-type stars observed only in the NIR, which will
be annexed to the GOSC. 
In this poster paper we present preliminary results derived from a set of
different attributes extracted from the retrieved papers. 

\end{abstract}

\begin{resumen}
Estamos llevando a cabo una recopilaci\'on bibliogr\'afica de los estudios
espectrosc\'opicos en el cercano infrarrojo ($0.7-5.0$ $\mu$m) disponibles
para las estrellas del Cat\'alogo de Estrellas O Gal\'acticas (GOSC, Ma\'iz
Apell\'aniz et al. 2004).
Esta recopilaci\'on nos permite cuantificar de forma precisa el grado de
conocimiento acerca de la informaci\'on espectral en el cercano infrarrojo
para las estrellas del GOSC, tales como banda de cobertura, resoluci\'on
espectral, medidas de anchos equivalentes, etc.
Este proyecto tiene un pr\'oximo paso definido que es el desarrollo de un
nuevo cat\'alogo de estrellas de tipo O observadas s\'olo en el cercano
infrarrojo, el cual ser\'a anexado al GOSC. 
En este trabajo presentamos resultados preliminares de un conjunto de
atributos extra\'idos de las publicaciones recolectadas.
\end{resumen}

\section{Motivation}

\noindent There are many open questions about the spiral structure and stellar
distribution in the Milky Way (MW). 
While the structure outlined by star forming regions (SFR) and molecular
clouds is relatively well known, it is difficult to be drawn using only
massive stars and young open clusters.
In spite of the progress made, the spiral structure beyond 2~kpc from the Sun,
as well as that on the far side of the Galaxy, is poorly determined.  
L\'epine et al. (2011) show how dramatic is this situation.
Their Figures 7 and 9 plot the distribution of CS and maser sources (tracers
of SFRs) in the Galactic Plane, in contrast with stellar optical/NIR tracers
like Cepheids and young open clusters. 
Comparing these two distributions, we can infer that about 90\% of the spiral
structure as traced by the youngest stellar populations is still completely
unknown.
 
On the other hand, the knowledge of the number and distribution of massive
stars is intrinsically important because these stars play a crucial role in
the dynamic and chemical evolution of the MW. 

Modern deep NIR sky surveys (e.g. VVV, Minniti et al. 2010; UKIDSS, Hewett et
al. 2006) are opening a new window of galactic explorations.
We now have the opportunity to discover an abundant population of hidden
massive stars from the stellar candidates selected from those surveys. 
Thus, NIR spectroscopy, through spectral classification, is an indispensable
tool allowing us to determine the nature of these candidates.

At present, the primary source of knowledge about massive stars (with 99\%
completeness at $B < 8$), is the {\it Galactic O Star
Catalog} (GOSC, Ma\'iz Apell\'aniz et al. 2004). 
The GOSC collects information for the optically brightest galactic 
O-type stars (370 objects), providing coordinates, spectral types, 
optical and near-infrared photometry, and other useful information.
Sota et al. (2008) presented the second version of GOSC, which seeks to extend
the catalog to $B < 14$.
Moreover, we are part of the {\it Galactic O-Stars Spectroscopic Survey}
(GOSSS, Ma\'iz Apell\'aniz et al. 2011), this all-sky intermediate-resolution 
spectroscopic optical survey of all O-type stars, is designed to revolutionize
the spectral classification system through high-quality and homogeneous
spectroscopic observations of more than one thousand O-type stars. 
Based on GOSSS data, Sota et al. (2011) re-discussed the spectral
classification system and presented a new atlas of O Stars. 

In the NIR regime, the general spectroscopic classification system for
O-type stars is based on a few atlases (e.g. Hanson et al. 1996, 2005). 
These atlases constitute the tool to perform a Morgan-Keenan (MK) process
of spectral classification in the NIR regime in a similar way as in the
optical. 
Unfortunately, the quality and incompleteness of the published NIR
spectroscopic atlases is not enough to reproduce the MK system in this regime, 
and many ambiguities are detected. 

We are planning a new NIR spectroscopic survey in order to establish the
basis of a spectroscopic classification system for O-type stars in NIR
comparable in quality and number of standards to that performed in the
optical by Sota et al. (2011).
As a first step, we are now retrieving bibliographic spectroscopic resources
for the GOSC sources in the NIR domain, which allow us to
quantify the precise degree of knowledge about NIR spectral information
for GOSC entries, like band coverage, spectral resolution, equivalent-width
measurements, etc.
This bibliographic compilation has a clear next step toward the development
of a new catalog of O-type stars observed only in the NIR (GOSC-IR), which
will be annexed to the GOSC.  

In this poster paper we present preliminary results derived from a set of
different attributes extracted from the retrieved papers.

\begin{figure}[!ht]
  \centering
  \includegraphics[width=0.45\textwidth]{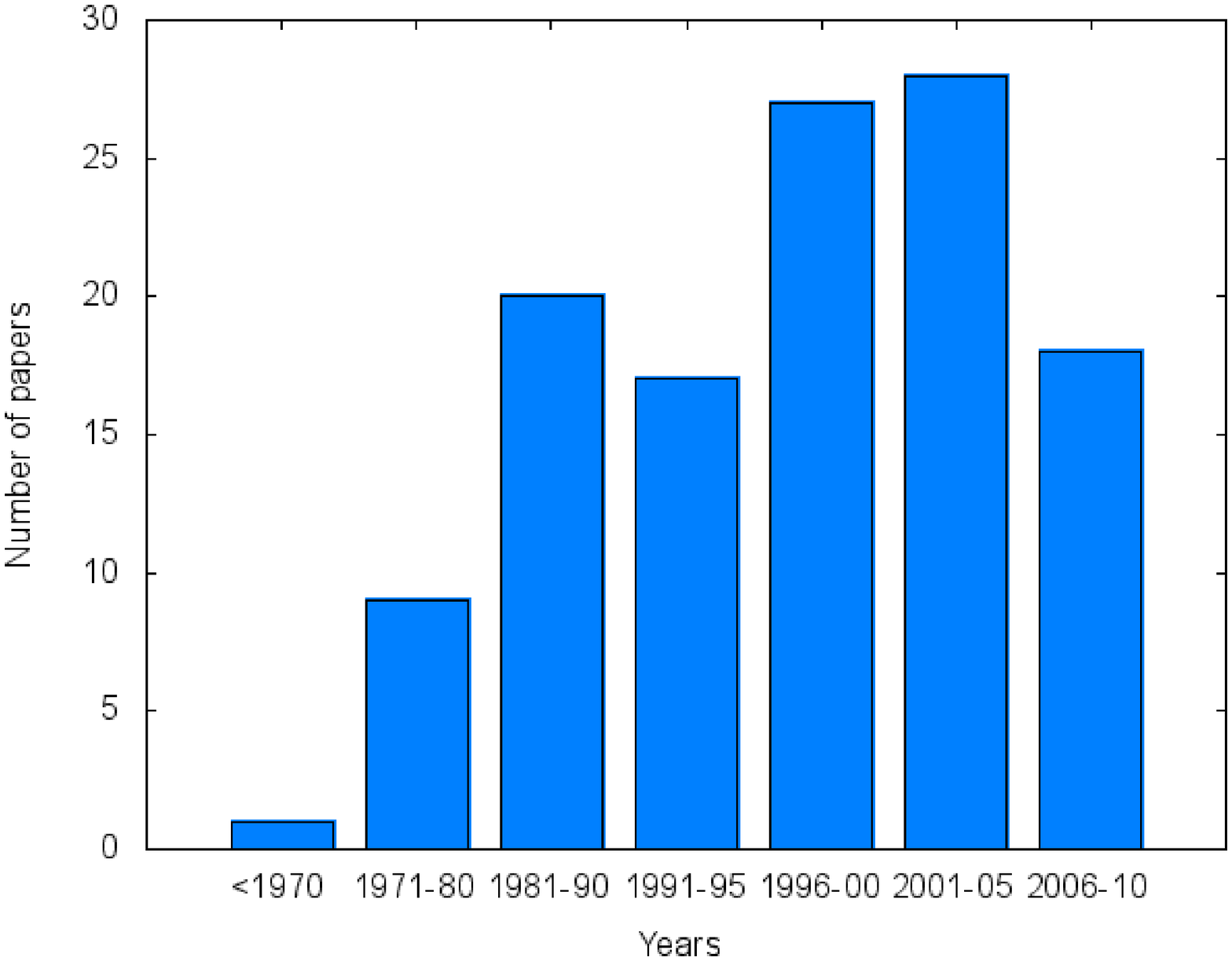}
  \includegraphics[width=0.45\textwidth]{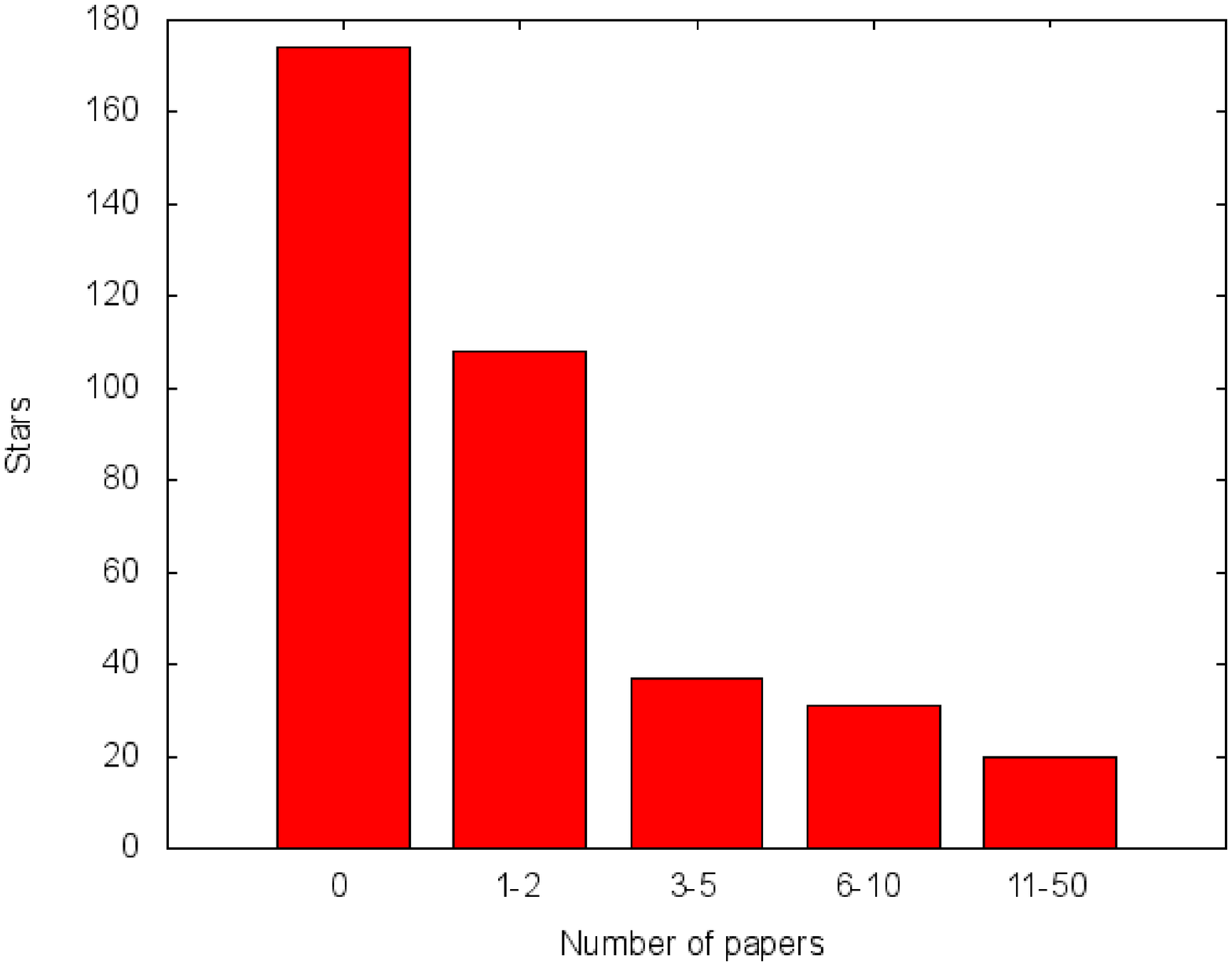}
  \caption{
    {\it Left:} Distribution of the number of published papers with NIR
    spectroscopy for O-type stars by year interval. 
    The first published paper showing an $I$-band photographic spectrum of an
    O-type star corresponds to {\it Spectre infrarouges de quelques \'etoiles
    des premiers types entre 6500 et 8800 A} by Andrillat \& Houziaux (1967).
    {\it Right:} Distribution of the number of O-type stars by number of
    papers. Note that almost 50\% of GOSC sources have no published NIR
    spectroscopic information.}
  \label{fig1}
\end{figure}

\section{The bibliographic catalog}

We retrieved the spectroscopic bibliographic information for 370 GOSCv1
sources using the NASA {\it Astronomical Data System} (ADS). 
We deployed different strategies to find papers which contain spectroscopic
observations (from X-rays to IR) of GOSC entries.
The total number of bibliographic entries counted is 28229 (including part of
year 2010), from which 837 entries are related to NIR spectroscopy for 196
stars (Figure~\ref{fig1}).  
These entries correspond to 123 independent papers. 
The first obvious result is that almost 50\% of the GOSC stars have not
published spectroscopic observation beyond 0.7~$\mu$m.
We must say that the definition of the NIR regime can be a bit tricky. 
We decided to include as a NIR spectroscopic observation those spectra
obtained in the spectral range 7000~$\AA$ to 5~$\mu$m.
Thus, the NIR lower limit can include many {\it optical} papers, as the $I$ and
$Z$ bands are accessible using CCD detectors or photographic plates.

From the bibliographic compilation we defined a number of {\it attributes} to be
developed in a catalog that lists the NIR spectroscopic information for GOSC
entries. The selected attributes are: {\tt BAND}, {\tt SPECTRAL RESOLUTION},
{\tt SPECTRAL BANDWIDTH}, {\tt SCIENCE CLASS}, {\tt FLUX}, {\tt PLOT}, {\tt
  DIGITAL DATA}, {\tt EQUIVALENT WIDTH}, {\tt RADIAL VELOCITY}, {\tt WIDTH
  PROFILE}. The {\tt SCIENCE CLASS} attribute refers to the primary purpose
for which the spectrum was obtained, i.e., to study the star itself, the ISM
or the nebular surrounding. 

The catalog will be published following Virtual Observatory standards in the GOSC
web page: {\bf http://ssg.iaa.es/en/content/gosc-v2-query/}  

\begin{figure}[!ht]
  \centering
  \includegraphics[width=0.45\textwidth]{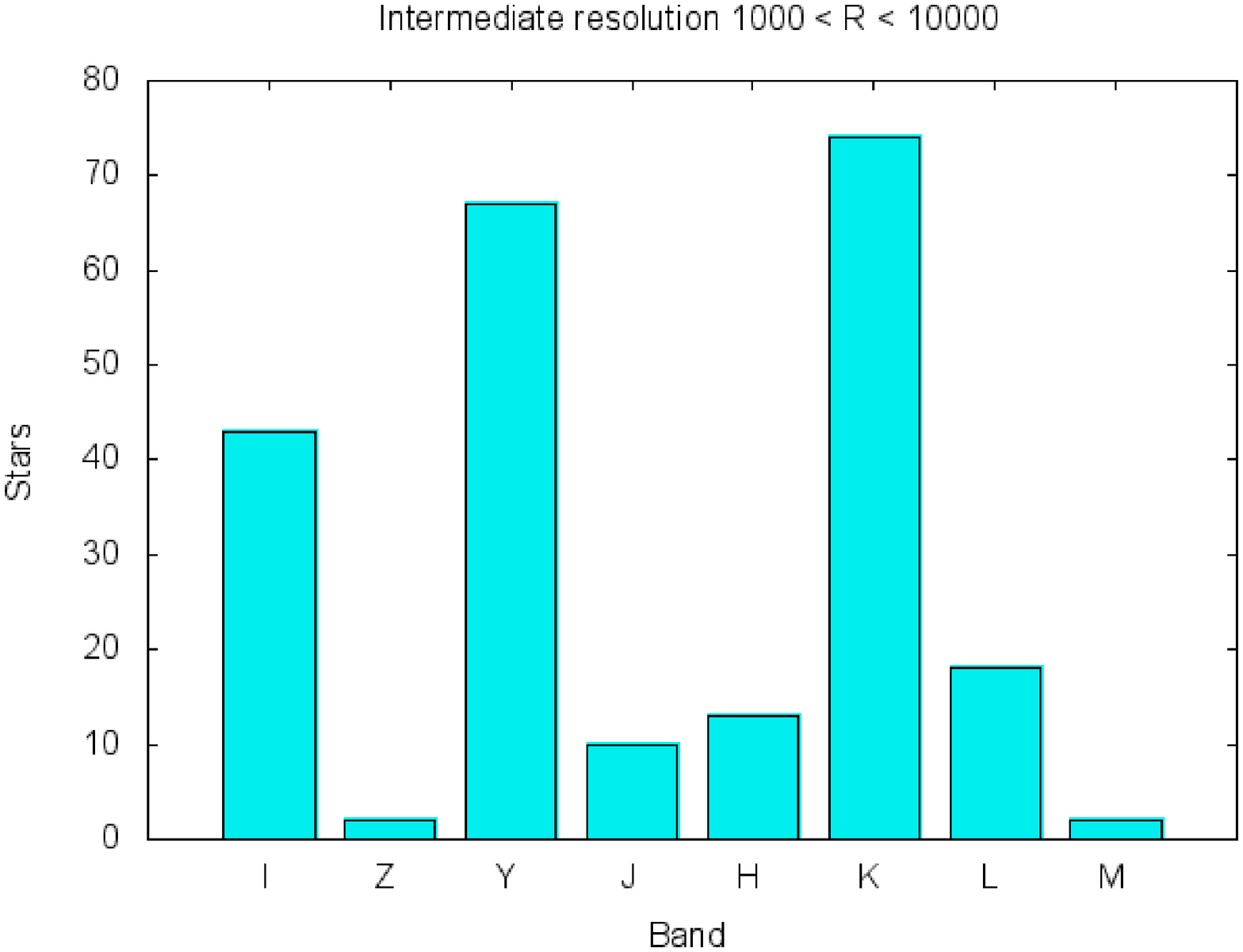}
  \includegraphics[width=0.45\textwidth]{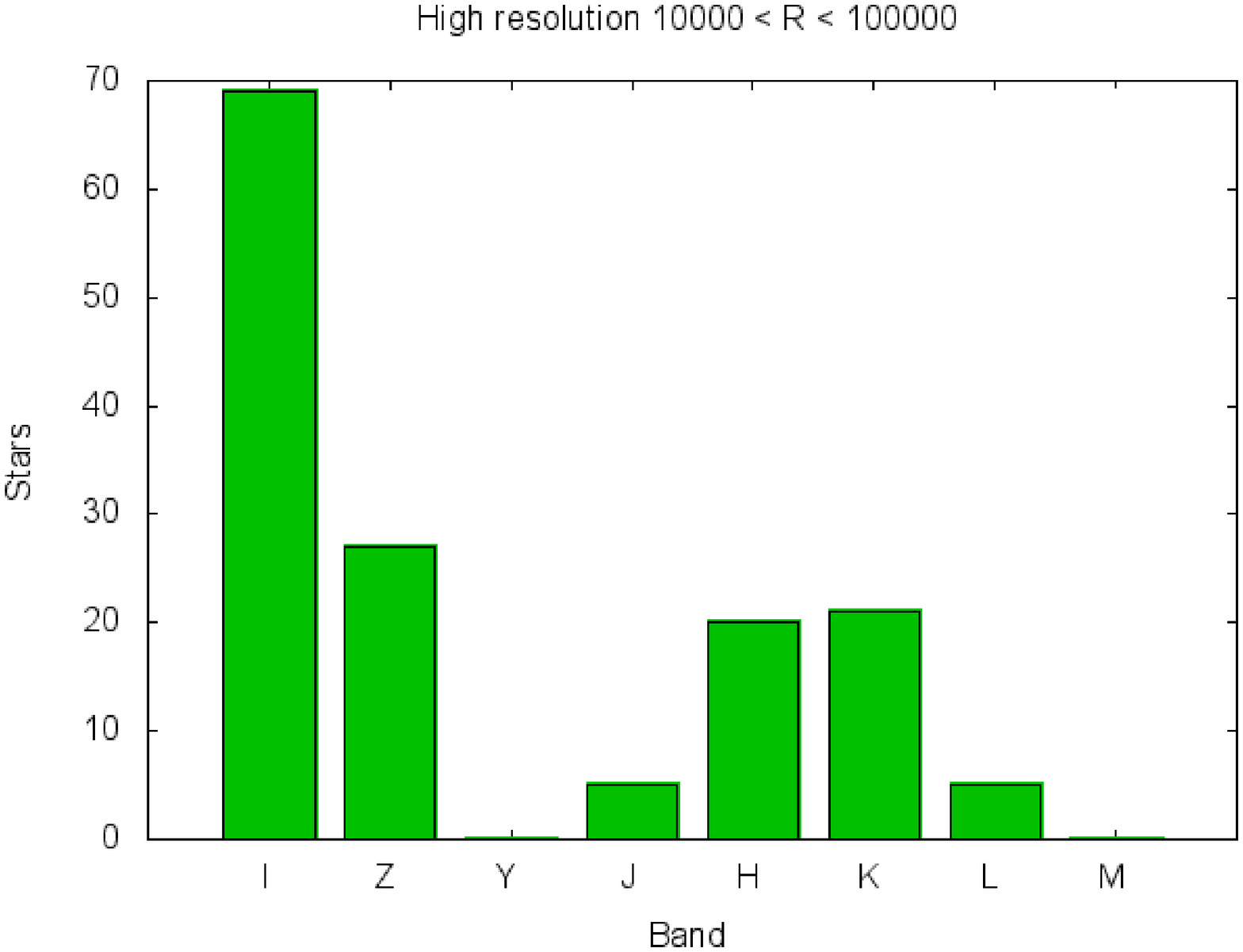}
  \caption{Number of stars observed in each NIR band at intermediate and high
    spectroscopic resolutions. 
    The distributions clearly show how inhomogeneous is the knowledge of the
    O-type stars in the different spectral bands. 
    This picture is even worst for specific spectral types, like O2, O3 or O6,
    which have never been observed at some resolutions. We note that there are
    only two observations of O-type stars at very-high spectral resolution (R
    > 100000) in the NIR.} 
  \label{fig2}
\end{figure}

Figure~\ref{fig2} shows some examples of the information which can be
extracted from the catalog.

At the moment, we have collected 663 different NIR spectroscopic observations
of 166 GOSC stars from 62 papers so far reviewed. Some interesting results are
already emerging. We find that radial-velocity and rotational broadening
studies are very scarce in the NIR, as well as ISM studies beyond 1\,$\mu$m. 
Also, most of the digital spectra available for researchers in astronomical
databases come from just four papers.
\medskip

\noindent {\bf Acknowledgments.}
\medskip

We wish to acknowledge to the referee, Dr. Margaret M. Hanson, for constructive
comments and suggestions which helped to improve the paper.
We thank financial support from DIULS PR09101, the {\it Facultad de Ciencias}
ULS, and DGAE-ULS. This research has made use of NASA's Astrophysics Data
System Bibliographic Services and SIMBAD database operated at CDS, Strasbourg,
France. 

\begin{referencias}
\reference Andrillat, Y.\& Houzieaux, L. 1967, J. Obs., 50, 107
\reference Hanson, M.M. et al. 1996, ApJS, 107, 281
\reference Hanson, M.M. et al. 2005, ApJS, 161, 154
\reference Hewett, P. et al. 2006, MNRAS, 367, 454
\reference L\'epine, J.R.D., Roman-L\'opes, A., Abraham, Z., Junqueira,
T.C. and Mishurov, Yu.N. 2011, MNRAS, 414, 138 
\reference Ma\'iz Apell\'aniz, J. et al. 2004, ApJS, 
\reference Ma\'iz Apell\'aniz, J. et al. 2011, {\it Highlights in Spanish
  Astrophysics VI}, Proceedings of the IX Scientific Meeting of the Spanish
Astronomical Society (SEA), held in Madrid, September 13 - 17, 2010, Eds.:
M. R. Zapatero Osorio et al., 467 
\reference Minniti, D. et al. 2010, NewA, 15, 433
\reference Sota, A. et al. 2008, RMxAC, 33, 56
\reference Sota, A. et al. 2011, ApJS, 193, 24 
\end{referencias}

\end{document}